\documentclass[conference]{IEEEtran}
\IEEEoverridecommandlockouts
\usepackage{cite}
\usepackage{url}
\usepackage{amsmath,amssymb,amsfonts}
\usepackage{algorithmic}
\usepackage{graphicx}
\usepackage{textcomp}
\usepackage{xcolor}
\def\BibTeX{{\rm B\kern-.05em{\sc i\kern-.025em b}\kern-.08em
    T\kern-.1667em\lower.7ex\hbox{E}\kern-.125emX}}
\begin{document}

\title{Secure Command, Control and Communications Systems (C3) for Army UxVs}

\author{\IEEEauthorblockN{1\textsuperscript{st} T.Rebolo}
\textit{Military Academy}\\
Lisbon, Portugal \\
rebolo.tn@academiamilitar.pt
\and
\IEEEauthorblockN{2\textsuperscript{nd} A.Grilo}
\textit{INESC INOV, Instituto Superior Técnico}\\
Lisbon, Portugal \\
antonio.grilo@inov.pt
\and
\IEEEauthorblockN{3\textsuperscript{rd} C.Ribeiro}
\textit{INESC INOV, Instituto Superior Técnico}\\
Lisbon, Portugal \\
ribeiro.cn@gmail.com}

\maketitle

\begin{abstract}
Unmanned Vehicles (UxVs) are increasingly used in modern military operations for reconnaissance, surveillance, and strike missions, enhancing situational awareness while reducing risk to personnel. Their affordability and rapid deployment have encouraged the adoption of commercial solutions. However, many rely on insecure protocols such as MAVLink, which lack authentication and encryption mechanisms. This paper designed, implemented, and evaluated a new secure command-and-control architecture that ensures confidentiality, integrity, and authentication (CIA) while supporting real-time control delegation between Ground Control Stations (GCSs). The proposed solution, named New Command and Control System (NC2S), enforces a zero-trust model integrating hierarchical credential-based privileges to regulate access and control among Tactical Commanders (TC), GCSs, and UxVs. It employs mutual Transport Layer Security (mTLS) with Elliptic Curve Digital Signature Algorithm (ECDSA) certificates and Elliptic Curve Diffie–Hellman (ECDH) key exchange, while message integrity is ensured through Hash-based Message Authentication Codes (HMAC). Multiple lightweight protocols were developed for credential management, key renewal, and control handover. The NC2S prototype was experimentally validated over Wi-Fi and Rohde \& Schwarz HR-5000H tactical radios. Results showed that HR-5000H links introduce latencies roughly two orders of magnitude higher than broadband technologies (e.g., Wi-Fi or 5G\&Beyond technologies) but are still able to maintain stable communication with minimal message loss, making them suitable for the NC2S links among TC terminals and GCSs.  
\end{abstract}


\section{Introduction}

Unmanned Vehicles (UxVs) have become indispensable assets in contemporary defense operations, enhancing reconnaissance, surveillance, and combat capabilities while minimizing the exposure of personnel to hazardous environments~\cite{Hossain,marques2013veiculos,ferreira_2012}. Beyond their operational relevance, the growing presence of commercial UxV platforms in military environments has made their integration into Command, Control, and Communications (C3) systems increasingly important. Commercial UxVs offer low acquisition cost, rapid procurement, and high availability, enabling armed forces to scale UxV fleets without the long development cycles typical of military-grade platforms. This affordability facilitates large-quantity deployment, supports experimentation and training, and accelerates modernization programs. However, as their operational use expands, the need for \textit{secure, interoperable, and resilient communication architectures} has become increasingly critical~\cite{hussain2014c4i}.

Current military C3 systems, such as the Portuguese Army’s \textit{Battlefield Management System (BMS)} and \textit{Dismounted Soldier System (DSS)}, provide tactical coordination and a Common Operational Picture (COP)~\cite{sequeira2020bms,DSS}. Nevertheless, they lack the capability to integrate and manage \textit{secure control of commercial UxVs}, particularly regarding \textit{control handover} between Ground Control Stations (GCS). This gap restricts operational flexibility and poses serious security risks in scenarios requiring distributed or multi-domain UxV coordination.

Commercial communication protocols, including the widely adopted \textit{Micro Air Vehicle Link (MAVLink)}, have been considered for UxV control due to their simplicity and interoperability \cite{koubaa2019mavlink,Meier2009MAVLink}. However, these protocols were not designed for contested military environments and suffer from major vulnerabilities such as lack of authentication, encryption, and integrity verification. Such weaknesses expose systems to a range of \textit{cyber threats}, including eavesdropping, spoofing, replay, denial-of-service (DoS), and man-in-the-middle (MITM) attacks \cite{kwon2018,Buckle,fereidouni2023iot,s23136067}. These attacks can compromise mission data confidentiality, disrupt control links, or allow adversaries to hijack UxV operations.

From a communications standpoint, UxV C2 architectures must operate across both civilian and military radio technologies. While 5G technologies are not yet widespread at the tactical edge, current C2 systems continue to rely on legacy tactical radios, such as the Rohde \& Schwarz HR-5000H and the EID PRC-525. These radios already implement robust \textit{Communication and Transmission Security} (COMSEC/TRANSEC) measures, ensuring link-level confidentiality and resistance to interception. However, they do not enforce higher-level \textit{access-control policies}: if a GCS is captured or misbehaves—intentionally or not—the radio cannot prevent it from issuing unauthorized commands or violating the tactical commander's intent. Moreover, these radios operate under bandwidth and latency constraints that limit real-time control of UxVs ~\cite{RohdeSchwarz2020,dissertacao}. Consequently, beyond physical-layer protection, C2 systems must integrate cryptographic and policy-based mechanisms capable of binding operational authority to mission-specific credentials. 

To address these issues, this work proposes the \textit{New Command and Control System (NC2S)}, a secure, lightweight, and flexible architecture for UxV operations within military C3 environments. NC2S adopts a \textit{zero-trust security model} incorporating hierarchical, credential-based access control among Tactical Commanders (TCs), GCSs, and UxVs. It employs \textit{mutual Transport Layer Security (mTLS)} with \textit{Elliptic Curve Cryptography (ECC)}, including the \textit{Elliptic Curve Digital Signature Algorithm (ECDSA)} for authentication and \textit{Elliptic Curve Diffie–Hellman (ECDH)} for key exchange. The system also defines a family of lightweight protocols for credential management, key renewal, and control delegation, ensuring mission continuity even under constrained communication channels.

The proposed NC2S framework was experimentally validated using both Wi-Fi and \textit{Rohde \& Schwarz HR-5000H} tactical radios, measuring connection establishment time, key renewal latency, and control handover efficiency. Results demonstrated that while tactical radios introduce significantly higher latency—approximately two orders of magnitude greater than Wi-Fi—they maintain stable and reliable communication suitable for command-node coordination. Conversely, broadband technologies (e.g., Wi-Fi for short-range communication or 5G for long-range communication) remain the preferred medium for high-frequency UxV control where throughput is critical.

\noindent\textbf{This work presents the following contributions relative to the state of the art:}

\begin{itemize}
    \item \textbf{NC2S Architecture:} A zero-trust C2 framework that separates \textit{long-term identity}, established through \textbf{X.509 ECDSA certificates}, from \textit{mission-scoped authorization}, enforced through \textbf{TC1-signed credentials} and a hierarchical \textbf{capacity policy} regulating per-link operations at runtime.

    \item \textbf{Lightweight Protocol Suite:} A set of compact and efficient protocols for registration, session-key establishment and renewal, credential and certificate renewal, and revocation, providing strong authentication and integrity even over bandwidth-constrained tactical radios.

    \item \textbf{Secure Control Handover:} A novel handover mechanism supporting real-time \textbf{transfer of UxV control between GCSs} through credential revocation or capacity-string modification, ensuring that command delegation strictly follows the Tactical Commander’s policy and preventing unauthorized takeover.

    \item \textbf{COTS Integration:} Seamless compatibility with COTS systems via Python-based proxy modules that secure and authenticate MAVLink traffic without modifying Mission Planner or ArduPilot software.

    \item \textbf{Experimental Validation:} A full prototype and performance assessment over \textbf{Wi-Fi and Rohde \& Schwarz HR-5000H} radios, demonstrating stable, authenticated communication with manageable latency and minimal message loss, confirming NC2S’s practicality for real military C4I environments.
\end{itemize}

The paper is organized into eight sections: Section II reviews related work, Section III presents the NC2S architecture, Section IV details the protocols, Section V covers the threat analysis, Section VI compares usability, Section VII reports the experimental validation, and Section VIII the paper and provides relevant paths for future work.

\section{Related Work}
\label{sec:related-work}

This section reviews prior work that informed the design space for secure C3 over UxVs.  
The literature clusters into three main strands: packet protection at the protocol layer, key establishment and management, and authentication with control-handover mechanisms.  
Together, these strands reveal recurring trade-offs among cryptographic assurance, latency, scalability, and field practicality over constrained links.

\subsection{Packet Protection and Cryptographic Protocols}
Prior works have proposed ways to implement per-packet confidentiality and integrity in MAVLink.

Allouch \textit{et al.} evaluated AES-CTR, AES-CBC, RC4, and ChaCha20 integrated into MAVLink under ArduPilot SITL, showing that ChaCha20 preserved near-baseline packet rate with low CPU and memory cost, while AES-CBC and RC4 were the most resource-intensive~\cite{allouch2019mavsec}.  Tüfekci \textit{et al.} advanced to AEAD, benchmarking ChaCha20-Poly1305, AES-GCM-SIV, AES-OCB3, and AES-CCM on a Raspberry Pi~4 with Pixhawk. ChaCha20-Poly1305 achieved the fastest performance, AES-OCB3 minimized CPU usage, and AES-CCM minimized memory, at the expense of increased per-packet work compared to plaintext MAVLink~\cite{Tufekci2024}.   

In conclusion, packet protection on commercial solutions is feasible with modern ciphers and AEAD.  
The gains in confidentiality and integrity come with measurable compute and airtime overhead that must be budgeted against the constraints of tactical radios.

\subsection{Key Establishment and Management}
Mechanisms for key creation, refreshing, and distribution under mobile and intermittent connectivity are especially relevant for the implementation of a secure tactical edge.

Ismael \textit{et al.} generated session keys using a Chebyshev chaotic map and encrypted the communication with the HIGHT block cipher, achieving fast key generation with minimal CPU and memory usage~\cite{Ismael2021}.  Hashmi and Munir combined ECDH for session setup with ChaCha20 for data encryption, obtaining secure keys and low computational cost in SITL~\cite{Hashmi2023}.  Bae \textit{et al.} introduced “saveless” hop-distance group keys using hash chains, which reduced storage and communication overhead while supporting mobility, although with more complex rekeying when the topology changed~\cite{bae2019authentication}.  Wen \textit{et al.} generate session keys via ECDH and protect these, along with credentials, by binding them to users via a biometric fuzzy extractor and PUF, granting stronger security than other solutions~\cite{wen2024secure}.  

Kwon \textit{et al.} presents a system where private keys of both the UAV and ZSP are computed via an ECC-based process that binds elliptic curve parameters, identification parameters, and a pre-shared master key. The session key is then derived in an ECDH-based algorithm. The results demonstrated low overhead, fast computation, and strong security~\cite{Kwon2022SecureHandover}.  Lin \textit{et al.} separated Master/Refresh from Session/Auth keys to improve throughput and ensure forward secrecy through refresh key rotation~\cite{Li2022CSECMAS}.  Khalid \textit{et al.} applied an AES–RSA hybrid system to reduce communication bits and setup time for constrained nodes. Here, the session keys are derived from a shared secret combined with nonces~\cite{HOOPOE}. Semal \textit{et al.} achieved two-round group key agreement in a certificateless setting using bilinear pairings, trading higher computation for scalability without classic PKI~\cite{CLPKC}.  

To sum up, no single keying strategy dominates across mobility, compromise resistance, revocation, and renewal practicality. Designs reducing infrastructure dependence often raise computation or storage demands, whereas PKI-based solutions simplify assurance properties at the cost of field logistics.

\subsection{Authentication and Control Handover}

Mechanisms for UxV control handover and privilege reconfiguration can provide the commander the advantage of tactical flexibility. On the other hand, authentication mechanisms must be in place to prevent or mitigate intrusion attempts, as well as to block compromised or captured nodes.

Ko \textit{et al.} combined ECDSA certificates with ECDH and HMAC.  Formal analyses were positive, and Drone-to-Drone links ran efficiently, but Drone-to-GCS authentication around 213~ms can be costly when handovers are frequent~\cite{ko2021drone}.  Mahmoud \textit{et al.} demonstrated a global PKI with mutual TLS (mTLS) and license-linked aircraft certificates—feasible yet sensitive to single-anchor availability and revocation bottlenecks~\cite{trust_prov}.  

Alternatives aim to reduce infrastructure round-trips.  
Ayati \textit{et al.} explored Kerberos-style tickets for multi-GCS delegation, improving scalability in simulation while assuming continuous reachability to authority servers~\cite{ayati2022secure}.  Bae \textit{et al.} \cite{bae2019authentication} in the same paper cited before, presented an authentication system that is based in the hashing of a pre-shared key for the GCS-UAV link and in digital certificates in the inter-GCS links. The handover is granted by the sending of a new session key to both the UAV and the new in-control GCS. Although lightweight, the system relies on secure pre-flight setup. Wang \textit{et al.} \cite{Handoverkey} proposed an LTE-oriented handover key-management scheme in which neighboring Ground Relay Stations (X2) or the MME (S1/inter-MME) distribute root-key material so each handover derives fresh Access-Stratum keys with only ~2880 bits of signaling, yielding fast computation and key separation against compromised relays but remaining LTE-specific. 

Some solutions reduce complexity by using lightweight trust anchors instead of full PKI.

Wen \textit{et al.} and Kwon \textit{et al.}, both previously discussed, also reduced system complexity by employing trust anchors. Wen \textit{et al.} used a Registration Authority as the anchor, efficiently achieving mutual authentication with only two lightweight messages but dependent on the RA’s secrecy and availability~\cite{wen2024secure}. Kwon \textit{et al.} instead leveraged neighboring ZSPs as relays and trust anchors, providing low measured cost and role-based access control under a deployed ZSP infrastructure~\cite{Kwon2022SecureHandover}. Choe \textit{et al.} proposed an ECC-based scheme under a certificate authority (CA) trust anchor, adding ultra-wideband (UWB) timestamps for location binding, improving resilience while reintroducing a single trust anchor~\cite{ECC-Based}.  

Operationally simple models also appear. Filho \textit{et al.} preloaded per-GCS keys and an allow-list so that the UxV could authenticate a new GCS during handover—efficient but inflexible for dynamic additions~\cite{Filho2023SecureHandover}.  

Human-in-the-loop MOMU concepts ensured deliberate transfer but introduced approval delays and teamwork dependence~\cite{Fern2018MOMU}.  Decentralized designs using blockchain or DAGs removed single points of failure and provided tamper-evident logs, but increased energy, storage, and latency~\cite{li2023blockchain, Blockchain-Based}.  

In conclusion, PKI-centric handovers provide strong guarantees but face revocation and availability bottlenecks.  Ticket-based and hop-by-hop methods reduce latency and infrastructure dependence but assume reliable anchors or pre-flight provisioning.  
Pre-shared and operator-approved models are robust to connectivity loss yet restrict flexibility in dynamic missions.

\subsection{Overall Perspective}
Across these strands, the state of the art shows that packet-level protection is practical on commercial platforms. Key management must support renewal and revocation without heavy online dependence, and handover must authenticate both entity and authority under intermittent links.  
Persistent gaps include sensitivity to single trust anchors, delays due to revocation, and rigidity of pre-shared models. These observations motivate architectures that bind certificate-backed identity to mission-scoped, capacity-aware credentials and restrict on-air exchanges to the smallest set that still supports secure, low-latency delegation.

\section{NC2S - New Command \& Control System }

This chapter describes the NC2S architecture and functioning. All the proxy scripts, protocols flowcharts, GUI images, and testing architectures images can be consulted in \cite{rebolo2025nc2s}.

\begin{figure}
    \centering
    
    \includegraphics[width=1\linewidth]{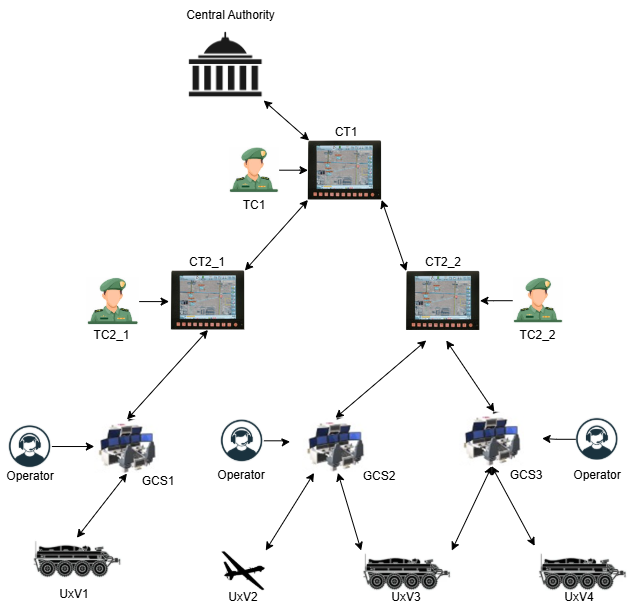}
    \caption{NC2S Architecture.}
    \label{fig:NC2S Architecture}
\end{figure}
 
\subsection{System Architecture and Entities}

The NC2S introduces a hierarchical, zero-trust architecture to integrate commercial unmanned vehicles into military C4I systems. At its core are the Certificate Authority (CA), Tactical Commanders (TC1 and TC2), Commander Terminals (CT1 and CT2), GCSs, and UxVs (Fig. \ref{fig:NC2S Architecture}). The CA acts as the trust anchor, generating and signing X.509 ECDSA certificates for every node and issuing certificate revocation lists (CertRLs). TC1 is the operational authority, defining the mission policy, requesting certificates and keys from the CA, creating and signing credentials granting nodes specific permissions, issuing credential revocation lists (CredRLs), and managing control delegation. The CT1, a Python application, provides TC1 with a graphical interface to manage certificates, credentials, connections and network mapping. It can connect to multiple CT2 and GCS nodes. TC2 and its terminal (CT2) provide an intermediate command layer under TC1’s control. The TC2 can relay credentials and initialize sessions with lower-level nodes but cannot request new certificates or issue credentials on its own.

A GCS interfaces directly with one or more UxVs. Each GCS runs Mission Planner to control the vehicle and a Python “proxy” script that implements NC2S message processing, cryptographic operations, and routing functions. The UxV side also contains a proxy interfacing with ArduPilot SITL or the real vehicle to unpack MAVLink messages and forward them to the proper destination.

\begin{figure}
    \centering
    \includegraphics[width=1\linewidth]{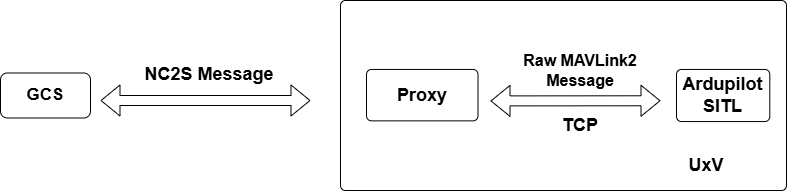}
    \caption{UxV Architecture.}
    \label{fig:UxV Architecture}
\end{figure}

\begin{figure}
    \centering
    \includegraphics[width=1\linewidth]{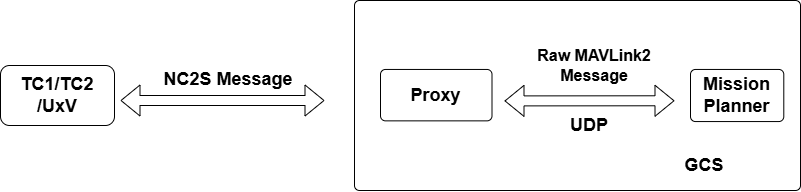}
    \caption{GCS Architecture.}
    \label{fig:GCS Architecture}
\end{figure}

This proxy-based integration is a key component for Commercial Off-The-Shelf (COTS) compatibility. The GCS proxy (Fig. \ref{fig:GCS Architecture}) intercepts raw MAVLink UDP packets from Mission Planner, encapsulates them in the secure NC2S message format, and forwards them to the authenticated UxV. In the reverse direction, it unwraps incoming NC2S messages to extract the raw MAVLink payload and forwards it to Mission Planner. Similarly, the UxV proxy (Fig. \ref{fig:UxV Architecture}) performs the same wrap/unwrap function, interfacing with ArduPilot SITL (or the flight controller) via a TCP or serial connection. This architecture effectively decouples the secure NC2S framework from the COTS control software, allowing secure communication without modifying the Mission Planner or ArduPilot source code.

\subsection{Assumptions and Threat Model}

NC2S operates under a zero‑trust philosophy: no node is implicitly trusted, and every message must be authenticated and authorized before it is acted upon. The security of the system is anchored in two offline components:
\begin{itemize}
    \item \textbf{A1 – Root of Trust (CA)}: the CA is assumed to be secure, uncompromised, and physically isolated. It issues X.509 ECDSA certificates and CertRLs on demand.
    \item \textbf{A2 – Policy and Authorization Anchor (TC1)}: the TC1 is the root of \textit{operational} trust and is assumed to be physically secure. While the CA validates a node's \textit{identity}, the TC1 validates its \textit{authority}. The system trusts the TC1 to define and enforce mission policies, issue and manage access credentials, maintain the validity of all security artifacts, and ensure the timely distribution of updated CredRL to regulate operational permissions dynamically.
    \item \textbf{A3 – Time Base}: nodes maintain bounded clock skew sufficient for timestamp‑based replay protection.
    \item \textbf{A4 - Transport-layer confidentiality}: The confidentiality is assumed to be provided by the transport layer, not by the NC2S application layer. For tactical-radio links (e.g., HR-5000H), COMSEC/TRANSEC provide confidentiality and contribute to availability. For WiFi/5G/IP links, an encrypted transport technology (e.g., WPA or a TLS tunnel) should be used. Optionally, an additional encryption key pair may be deployed for payload encryption. Its negotiation/provisioning is out of scope for this thesis.
\end{itemize}

Because NC2S is designed for contested networks, the adversary model is strong. Following a Dolev–Yao approach \cite{Dolev}, we assume the adversary can:
\begin{itemize}
    \item \textbf{T1 – Globally Observe}: passively capture traffic on any radio or IP link and store it for later analysis.
    \item \textbf{T2 – Actively Interfere}: inject, drop, reorder or delay packets and mount MITM attacks.
    \item \textbf{T3 – Capture Field Nodes}: physically compromise any deployed node except the offline CA and TC1, extracting keys, credentials and logs.
    \item \textbf{T4 – Misuse Credentials}: operate rogue nodes using valid but constrained credentials obtained through theft or insider compromise.
    \item \textbf{T5 – Abuse Revocation and Time}: delay delivery of CredRLs/CertRLs or desynchronize clocks to attempt replay.
    \item \textbf{T6 – Exploit Cryptographic Weaknesses}: exploit misconfigurations or implementation flaws. 
    However, standard primitives (ECDH, ECDSA, HMAC) are assumed secure.
\end{itemize}

These assumptions and threats motivate NC2S’s design choices.

\subsection{Message Format and Basic Elements}

Each NC2S message has the form

\begin{equation}
  m = (\mathsf{type}, \mathsf{payload}, T,\ \mathsf{HMAC}_K(\mathsf{type} \,\|\, \mathsf{payload} \,\|\, T)),   
\end{equation}

where:
\begin{itemize}
  \item $\mathsf{type}$ (1~byte) identifies the message semantics, such as encapsulated MAVLink data,
        key-renewal triggers, credential distribution, or CertRL/CredRL updates.
  \item $\mathsf{payload}$ is the data being transported, such as a raw MAVLink packet or a control instruction.
  \item $T$ (8~bytes) is the sender timestamp used for replay protection. Nodes maintain a sliding window and discard duplicates or messages outside a 400 ms window.
  \item $\mathsf{HMAC}_K(\cdot)$ is a 16-byte truncated HMAC-SHA256 computed over
        $(\mathsf{type}, \mathsf{payload}, T)$ using a \emph{directional} session key $K$
        derived during connection establishment. Assures message authentication and integrity.
\end{itemize}

\subsection{Credentials, Capacity Policy and CertRL}

NC2S separates long-term identity from operational authorization. Identities are established via CA-signed certificates, while \emph{what} each node is allowed to do is governed by TC1-signed credentials and a mission-specific capacity policy, with enforcement supported by the CredRLs.

For each authorized unidirectional link $C \rightarrow S$, TC1 issues a credential

\begin{equation}
      Cred_{C \rightarrow S}
  = (\mathsf{type}, PK_C, PK_S, \mathsf{cap}, t_{iss}, t_{exp})
  \; Sig_{SK_{\mathsf{TC1}}}(\cdot)
\end{equation}

with the following structure:
\begin{itemize}
  \item $\mathsf{type}$ (1~byte): encodes the relationship / link type
        (e.g., TC1--TC2, TCx--GCS, GCS--UxV).
  \item $PK_C$ (256~bits): ECDSA public key of the client node.
  \item $PK_S$ (256~bits): ECDSA public key of the server node.
  \item $\mathsf{cap}$ (variable length): capacity string describing the permitted operations.
  \item $t_{iss}$ (8~bytes): issuance timestamp.
  \item $t_{exp}$ (8~bytes): expiration timestamp.
  \item $Sig_{SK_{\mathsf{TC1}}}(\cdot)$: ECDSA-SHA256 signature computed by TC1 over all
        previous fields.
\end{itemize}

Credentials are \emph{directional} and \emph{mission-scoped}: they authorize a specific $C \rightarrow S$
flow to perform only the operations encoded in $\mathsf{cap}$, during $[t_{iss}, t_{exp}]$. They may be provisioned during mission setup or distributed using authenticated NC2S messages.

The capacity policy, inspired by NATO STANAG 4586 LOI concept, defines mission privileges through categories that are divided into hierarchical sections. The categories themselves are cumulative, this is a credential may combine multiple capabilities within its capacity string. The policy is implemented in a JSON mission-file mapping each capacity string to the message types a node may send or receive. During operation, a node validates every message by checking its type against the credential’s capacity and the local policy, ensuring strict, runtime authorization.

The Credential Revocation List (CredRL) is specific to NC2S and is used to invalidate previously issued
credentials. Its structure is

\begin{equation}
      \mathsf{CredRL} = (T_{\text{iss}}, \mathsf{HL}, T_{\text{exp}}, Sig_{SK_{\mathsf{TC1}}}(T_{\text{iss}} \,\|\, T_{\text{exp}} \,\|\, \mathsf{HL})),
\end{equation}

where:
\begin{itemize}
  \item $T_{\text{iss}}$ (8~bytes) is the creation timestamp;
  \item $T_{\text{exp}}$ (8~bytes) is the expiration timestamp (CredRL validity limit);
  \item $\mathsf{HL}$ is the hash list containing the identifiers (hashes) of revoked credentials;
  \item $Sig_{SK_{\mathsf{TC1}}}(\cdot)$ is an ECDSA over SHA-256 signature computed by TC1.
\end{itemize}


Because these credentials are unique to NC2S, it is useful to briefly outline how they are verified. When a node receives a credential: (i) verify that a valid (non-expired, correctly signed) CredRL is locally stored; (ii) check that the credential type $\mathsf{type}$ is consistent with the link context; (iii) verify the ECDSA-SHA256 signature using TC1's public key $PK_{\mathsf{TC1}}$; (iv) ensure the credential validity interval $[t_{iss}, t_{exp}]$ includes the current time; (v) confirm that $PK_C$ and $PK_S$ match the expected client and server public keys for this session; (vi) compute the hash of $Cred_{C \rightarrow S}$ and verify that it does not appear in the stored $\mathsf{HL}$ of the current CredRL. This verification ensures that only valid, authorized credentials are accepted and that revoked credentials cannot be reused.

\subsection{Certificates and CertRL}

NC2S uses X.509 certificates to establish long‑term identities for all nodes. The CA generates a unique prime256v1 (256‑bit) ECDSA key pair for every node and signs the corresponding certificate using its own ECDSA private key. Certificates bind a node’s common name (CN) and IP address to its public key and include timestamps for issuance and expiration. Nodes store the CA’s self‑signed certificate to verify signatures and ensure that the issuing authority is trusted.

The CertRL is issued by the CA when certificates must be revoked. It contains the serial numbers of all revoked certificates and, similar to the certificates themselves, includes both a creation timestamp and a validity period indicating the expiration date of that list.

To verify a certificate, a node checks the following: (i) if it has stored a valid CertRL; (ii) the certificate’s signature is valid under the CA’s public key; (iii) the certificate is within its validity period; (iv) the serial number is not listed in the stored CertRL; and (v) the certificate’s subject fields (CN and IP) match the expected peer identity. Only if all checks pass does the node accept the certificate.

\subsection{Mission Setup and Onboarding}

Before deployment, all nodes are “imprinted” in a secure, co‑located environment. TC1 defines IP addresses and port assignments, requests certificates and key pairs from the CA for every node, generates an initial empty CertRL and CredRL, and defines the mission capacity policy in a JSON file. Configuration files distributed to each node include the CA certificate, the node’s own certificate and private key, TC1’s public key, its server and client port assignments, empty revocation lists and the mission policy. TC1 retains a master list of all certificates and credentials for management. This onboarding ensures that each node has the cryptographic material and policy context needed to join the network and that no field provisioning is required.

\subsection{Commander Interface and Management}
\begin{figure*}[!t]
    \centering
    \includegraphics[width=0.7\linewidth]{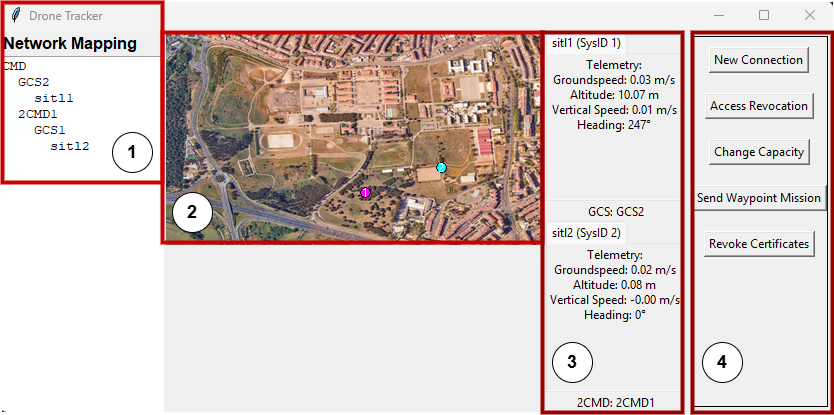}
    \caption{CT1 GUI.}
    \label{fig:CT1_GUI}
\end{figure*}

The TC1 and TC2 interact with the NC2S framework through a Python-based graphical user interface (GUI). As shown in Fig. \ref{fig:CT1_GUI}, the CT1 interface provides the commander with integrated monitoring and control. The CT2 interface is visually similar, providing the same network map and telemetry displays, but omits the command buttons (e.g., "Access Revocation", "New Connection"), reflecting its subordinate role.

The interface displays a real-time network map, populated by the Network Mapping Protocol (Sec. \ref{protocols}), showing all connected nodes. It also displays telemetry data from UxVs and provides command buttons to execute the core security protocols. For example, the "New Connection", "Access Revocation", and "Change Capacity" buttons directly initiate the Connection Establishment, Credential Revocation, and Control Delegation protocols, respectively. This GUI provides the essential human-in-the-loop mechanism for mission management, translating the commander's intent into secure, authenticated actions within the network.

\section{NC2S Protocols}\label{protocols}

This section describes the NC2S protocols.

\subsection{Connection Establishment and Session Key Agreement}

When TC1 initiates a connection between two nodes, it selects the client (initiator) and server, defines the server’s IP, mTLS port and UDP port, sets a credential lifetime and capacity string, and signs a credential. If TC1 cannot reach the destination directly, it encapsulates the credential and connection parameters into message types 0x05, 0x08 or 0x09 and routes them via CT2 and GCS proxies.

\begin{enumerate}
    \item \textbf{Handshake:} The client initiates an mTLS handshake to the server’s designated port. If the server is unreachable, the client retries up to three times with 20-second intervals.
    
    \item \textbf{Certificate Exchange:} Client and server present their ECDSA certificates and each verifies the peer’s certificate as explained before. After the certificates authentication the nodes negotiate the mTLS encryption keys.
    
    \item \textbf{Credential Transmission:} The client sends the signed credential to the server over the established TLS channel.
    
    \item \textbf{Credential Verification:} The server verifies the received credential.
    
    \item \textbf{Clock Synchronization:} The nodes perform a simple NTP exchange to estimate clock offset and set the baseline for timestamp validation.
    
    \item \textbf{Session Key Derivation:} The nodes generate a master secret via ECDH using their key pairs. Two session keys (one for each direction) are derived with an HKDF using SHA-256.
    
    \item \textbf{UDP Channel Setup:} The client tells the server which UDP port will be used for subsequent data exchange. Both nodes create the session list for that connection.
\end{enumerate}

Because this protocol uses mTLS, the confidentiality and integrity of the key exchange are ensured even over untrusted networks. Once the session is established, all traffic switches to UDP.

    

\subsection{Credential and Certificate Revocation}

NC2S supports two revocation scopes with distinct effects: \textbf{certificate revocation} (node eviction) and \textbf{credential revocation} (link/policy revocation). 

\textbf{Certificate revocation.} When TC1 wants to remove a node from the network (e.g., node capture), it requests the CA to revoke the node’s certificate. The CA issues a new CertRL, which the TC1 broadcasts. Upon receipt, each node: (i) verifies the CA signature, (ii) checks CertRL validity (timestamp/lifetime), (iii) ensures freshness (newer than stored), (iv) substitutes the older version with the received one and rebroadcasts it, (v) immediately terminates any session whose peer certificate serial appears in the list.

\textbf{Credential revocation.} When TC1 needs to shut down a \textit{link/permission} without evicting nodes, it issues a new \textit{CredRL} and also broadcasts it. Each node, when receiving a CredRL: (i) verifies TC1’s signature, (ii) checks validity, (iii) ensures freshness, (iv) substitutes the older version with the received one and rebroadcasts it, (v) closes any active session whose credential hash matches the CredRL hash list. 

By broadcasting the revocation lists across the network, this protocol increases the probability of all nodes receiving the latest revocation list, thereby maintaining consistent authorization status and preventing unauthorized access within the system.

NC2S supports two forms of control transfer between GCSs. The first approach addresses the scenario in which the control of a UxV must be transferred from one GCS (currently in control) to another GCS that is not yet connected to the UxV. Such a situation typically arises from tactical implications that also require the previous GCS to lose access to the UxV.

In this case, the TC1, through its CT1, creates a new CredRL that revokes the in-control GCS credential assigned to that UxV link and broadcasts the CertRL. After this, it initiates the session establishment protocol so the new GCS can initiate a new connection with the UxV with a credential that allows it to control that UxV.

The second approach addresses the situations where both GCSs are already connected to the same UxV. Here, one GCS currently holds control, and the TC1 intends to transfer control to the second GCS while allowing the first to retain a lower level of access. This approach eliminates the need for full reauthentication or credential revocation, thereby minimizing handover latency. The handover process is then based on the credential update mechanism where, using the “Change Capacity” function in CT1, TC1 generates new credentials similar to the current ones but modifies the capacity string field of each credential so the successor credential includes the Control permission, and the predecessor’s credential downgrades to a monitoring level. These credentials are then delivered via message types 0x13, 0x14 or 0x15, depending on the target, and are validated and stored by the receiving nodes. Since certificates and session keys remain valid, handover latency is significantly lower.

\subsection{Key, Credential and Certificate Renewal}

All security artifacts have explicit lifetimes. Each node runs routines to monitor expiry thresholds. When a session key approaches expiry, a client node sends a 0x11 key-renewal request message to the server and adds the server node  to a “key renewal” list. The server node, after receiving and verifying the 0x11 message, derives new keys from the existing ECDH secret using a freshly generated 4‑byte nonce, stores them alongside the old keys, responds with the nonce, and sets that client peer session to a “key renewal” state. Upon receiving the nonce, the client derives the new keys and replaces the old keys in its session state. Thus, the following messages sent to that server node will have the HMAC computed with the new keys. The server monitors incoming messages: if a message fails HMAC verification with the old key, it verifies if that peer session is in the “key renewal” state and recomputes the HMAC with the new key. Once a valid message is received under the new key, the server discards the old keys and returns to normal operation. This mechanism avoids transmitting key material across the network and prevents key exhaustion attacks.

Credential renewal follows a similar pattern. CT1 periodically checks the lifetimes of active credentials. When a credential is near expiration, it generates a new credential with the same parameters and routes it to the client and server nodes of that credential link. Certificate renewal is also handled by CT1, which periodically checks the lifetimes of active certificates and when near expiration, the CT1 requests fresh certificates from the CA and routes them to the destined nodes. CredRL and CertRL renewals are likewise handled by CT1 and CA, ensuring that revocation lists do not accumulate expired entries.

\subsection{Network Mapping}

To support situational awareness, GCSs periodically send 0x06 messages containing lists of connected UxVs. CT2 aggregates these into 0x10 messages for CT1, enabling a network map view on both CT1 and CT2 GUIs. 

\section{Threat and Usability Analysis}
This section provides an informal evaluation of the security properties of the proposed NC2S system
against a Dolev–Yao network adversary \cite{Dolev} and the common cyberattacks.

\subsection{Eavesdropping Attack (Confidentiality)}
Passive interception of radio or IP links cannot reveal information, as all traffic is encrypted at the transport layer (COMSEC/TRANSEC or TLS/WPA tunnels). Even full-channel observation yields no plaintext exposure.

\subsection{Message Tampering Attack (Integrity)}
Each packet includes an HMAC computed with direction-specific session keys derived via ECDH. Altered or forged packets fail verification and are discarded, ensuring integrity and origin authenticity.

\subsection{Impersonation Attack (Peer Spoofing)}
Certificate and credential verification during session establishment, together with per-packet HMAC, binds each message to an authenticated peer. Spoofing attempts without valid cryptographic material are rejected.

\subsection{Replay Attack}
Each message includes a timestamp validated within a bounded acceptance window. Out-of-window or duplicate timestamps are ignored, effectively preventing replayed messages.
 
\subsection{MITM Attack}
The combination of mutual certificate verification and per-packet HMAC prevents undetected alteration or session splicing. Any traffic not matching the expected authentication tags is dropped.

\subsection{Desynchronization Attack / Availability in Constrained Environments}
The NC2S implements bounded retries, authenticated renewal timers, and automatic re-initialization procedures that allow both peers to re-establish synchronized state. These mechanisms sustain availability even under intermittent or degraded tactical links such as narrowband radios.

\subsection{Privilege Escalation Attack}
Each credential embeds a capacity string defining authorized operations. Every command is verified against this field and the credential validity, denying actions beyond granted privileges or under revoked credentials. This enforces fine-grained access control and secure control handover.

\subsection{Key and Node Compromise Attack}
If a node is captured, its certificates and credentials are promptly revoked through CertRL/CredRL propagation. Short credential lifetimes, periodic key renewal, and direction-specific session keys confine the impact window of any leaked material.

\subsection{Usability Analysis}

Table~\ref{tab:comparison_nc2s_related} compares the proposed NC2S with selected systems reviewed in Section~\ref{sec:related-work}, considering the attack scenarios analyzed previously and key operational requirements. A $\checkmark$ indicates that the paper addresses or protects against the attack (or supports the feature), while “x” means it does not. 

\begin{table}[!t]
\centering

\renewcommand{\arraystretch}{1.2}
\setlength{\tabcolsep}{3pt}
\caption{Comparison between the proposed NC2S system and selected related works.}
\begin{tabular}{|p{5cm}|c|c|c|c|c|}
\hline
\textbf{Feature} &
~\cite{Handoverkey} &
~\cite{HOOPOE} &
~\cite{trust_prov} &
~\cite{Filho2023SecureHandover} &
\textbf{NC2S} \\
\hline
Eavesdropping & $\checkmark$ & $\checkmark$ & $\checkmark$ & $\checkmark$ & $\checkmark$ \\
\hline
Message Tampering & $\checkmark$ & $\checkmark$ & $\checkmark$ & $\checkmark$ & $\checkmark$ \\
\hline
Replay Attack & x & $\checkmark$ & $\checkmark$ & x & $\checkmark$ \\
\hline
Impersonation Attack & $\checkmark$ & $\checkmark$ & $\checkmark$ & $\checkmark$ & $\checkmark$ \\
\hline
MITM & $\checkmark$ & $\checkmark$ & $\checkmark$ & $\checkmark$ & $\checkmark$ \\
\hline
Desynchronization Attack / Availability in Constrained Environments & x & $\checkmark$ &$\checkmark$ & x & $\checkmark$ \\
\hline
Privilege Escalation & x & x & $\checkmark$ & x & $\checkmark$ \\
\hline
Node Capture Attack & $\checkmark$ & $\checkmark$ & $\checkmark$ & x & $\checkmark$ \\
\hline
UxV Control Handover & $\checkmark$ & $\checkmark$ & x & $\checkmark$ & $\checkmark$ \\
\hline
Integration with Commercial Systems  & x & x & x & $\checkmark$ & $\checkmark$ \\
\hline
Dynamic Node Join (New Nodes After Network Launch) & x & $\checkmark$ & $\checkmark$ & x & $\checkmark$ \\
\hline
Experimental Validation & x & x & $\checkmark$ & $\checkmark$ & $\checkmark$ \\
\hline
\end{tabular}

\label{tab:comparison_nc2s_related}
\end{table}

The NC2S stands out for enabling secure control handover through a hierarchical capacity policy that prevents privilege escalation. It was also experimentally validated with a working prototype, whereas most related systems remain theoretical or simulation-based. Moreover, NC2S is directly compatible with commercial solutions such as ArduPilot Mission Planner and UxVs using the MAVLink protocol.

\section{Experimental Validation}

\begin{figure}
    \centering
    
    \includegraphics[width=1\linewidth]{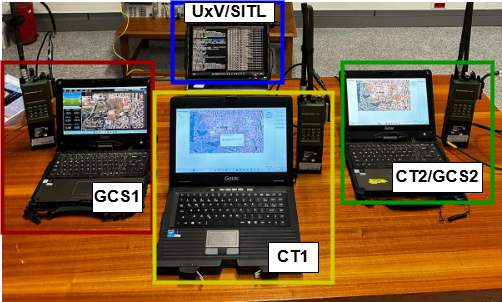}
    \caption{System Prototype Testing Environment.}
    \label{fig:Foto_testes}
\end{figure}

After designing and implementing the NC2S system, its performance and applicability were tested to evaluate its real-world behavior and validate its operational efficiency. The prototype, presented in Fig. \ref{fig:Foto_testes}, was composed of four nodes, each representing a system entity in the NC2S architecture:

\begin{itemize}
    \item \textbf{PC1 – TC1:} Getac (Windows), Intel i5-4210M~@~2.6~GHz, 8~GB~RAM, 256~GB~SSD.
    \item \textbf{PC2 – GCS1 and PC3 – TC2/GCS2:} Getac (Windows), Intel i5-6300~@~2.5~GHz, 8~GB~RAM, 256~GB~SSD.
    \item \textbf{PC4 – UxV:} Microsoft Surface~6~Pro (Windows), Intel~i7,~16~GB~RAM,~512~GB~SSD.
\end{itemize}

\begin{figure*}[!t]
    \centering
    
    \includegraphics[width=0.7\linewidth]{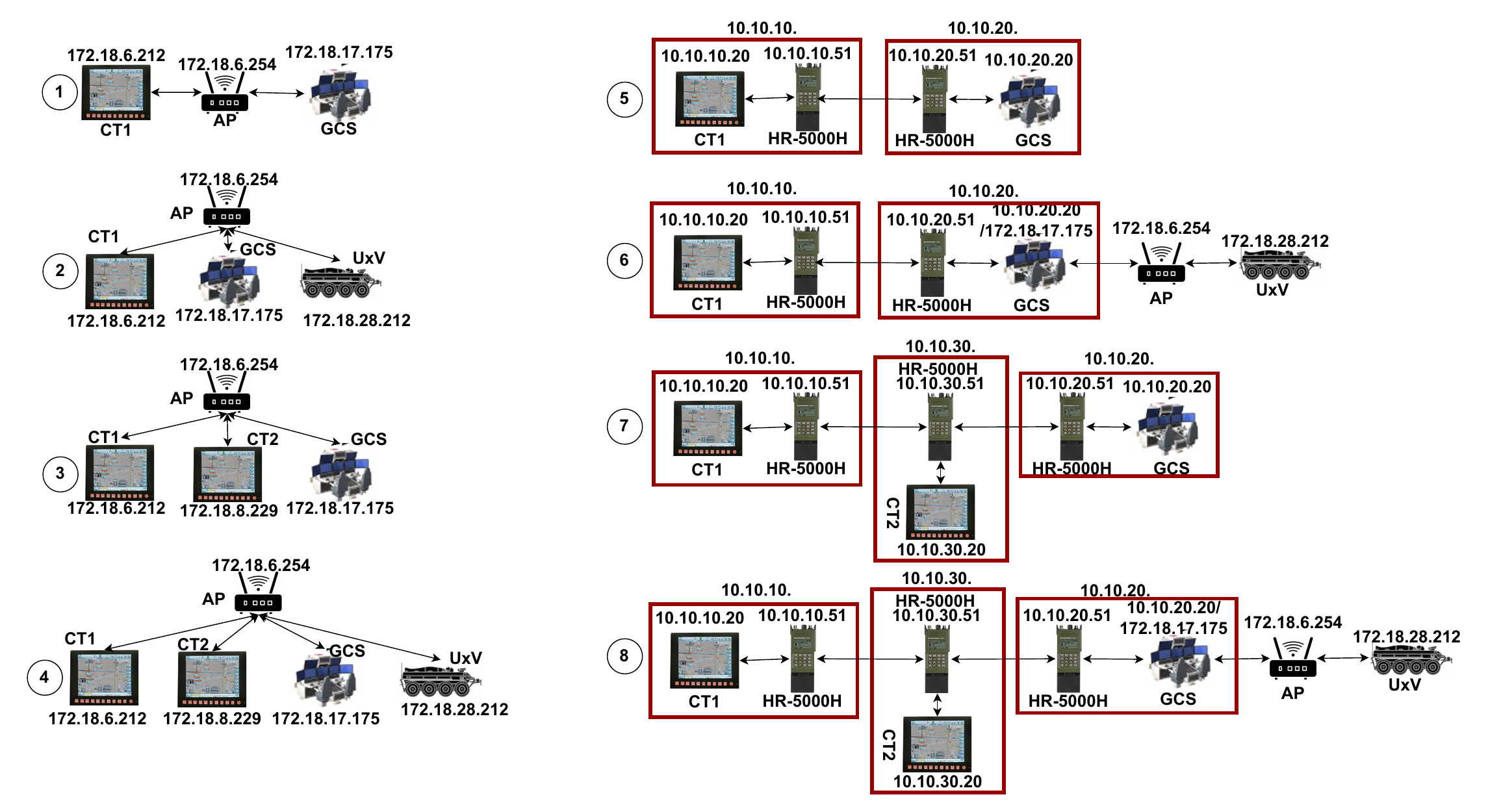}
    \caption{Connection Time, Key Renewal, System Reliability, Goodput Estimation and CPU Processing Time Testing Architecture.}
    \label{fig:Testing_Architecture}
\end{figure*}

Two communication setups were evaluated. For the broadband setup, Wi-Fi was used, and the system operated using an access point from the Army’s internal network, the Cisco~AIR-CAP1702I-E-K9 (\textit{802.11ac}) model. For the radio-based configuration, each HR-5000H radio was connected to a dedicated PC through Ethernet. The radios operated as network gateways, creating independent subnets per node. Each PC was configured with IP~\texttt{10.10.\textit{n}.20}, and its corresponding radio acted as the default gateway with IP~\texttt{10.10.\textit{n}.51}, where \textit{n} identifies the node. This design enabled complete logical isolation between links, mirroring field deployment conditions. These setups are displayed in Fig. \ref{fig:Testing_Architecture}.

The radios used were the Rohde \& Schwarz HR-5000H units operating with the SOVERON\textsuperscript{\textregistered}~WAVE~TNW50 Tactical Network Waveform in Robust Mode.  
The radio operated inside the 235-238.150 MHz band, divided into 64~channels spaced by~50~kHz, employing both COMSEC and TRANSEC in order to simulate a real battlefield implementation.

\subsection{Baseline Channel Characterization}

To establish reference conditions, iperf3 and ping tests were executed between PC1 and PC2 in both directions. The result of these tests can be visualized in Table \ref{tab:iperf_results_bidirectional}. Over Wi‑Fi, TCP throughput averaged 25–29 Mb/s, and UDP goodput reached up to 73 Mb/s for a 100 Mb/s offered load. Packet loss remained below 1\%, jitter was under 1.2 ms and the mean round‑trip time (RTT) measured via ping was around 5.5 ms. In contrast, the HR‑5000H radios sustained only 3.8–5.0 kb/s UDP throughput (inside the 10 Kb/s limit rated by the manufacteur) with RTTs between 839–968 ms and jitter ranging from 55 ms to 780 ms. TCP tests were inconclusive. One possible justification is that the high link latency caused repeated retransmission timeouts, leading to buffer saturation and session termination. These baseline results underscore the two orders of magnitude difference in bandwidth and latency between the wireless LAN and the tactical radio.

\begin{table*}[!t]
    \centering
    \caption{Mean \textit{iperf3} results for TCP and UDP tests between the nodes using Wi-Fi and HR-5000H radio links, measured in both directions (PC1–PC2 and PC2–PC1).}
    \begin{tabular}{|c|c|c|c|c|c|c|}
    \hline
     & \multicolumn{3}{c|}{\textbf{PC1–PC2}} & \multicolumn{3}{c|}{\textbf{PC2–PC1}} \\ \hline
     & \textbf{Throughput} & \textbf{Packet Loss} & \textbf{Jitter} & \textbf{Throughput} & \textbf{Packet Loss} & \textbf{Jitter} \\ \hline
    \multicolumn{7}{|c|}{\textbf{Wi-Fi}} \\ \hline
    TCP & 25.5~Mb/s & -- & -- & 28.8~Mb/s & -- & -- \\ \hline
    UDP test 100~Mb/s & 52.54~Mb/s & 13.158~\% & 0.1746~ms & 73.02~Mb/s & 7.14~\% & 0.1489~ms \\ \hline
    UDP test 50~Mb/s  & 47.34~Mb/s & 2.582~\%  & 0.6134~ms & 47.62~Mb/s & 4.156~\% & 0.783~ms \\ \hline
    UDP test 40~Mb/s  & 39.64~Mb/s & 0.554~\%  & 1.2108~ms & 38.34~Mb/s & 3.168~\% & 0.4502~ms \\ \hline
    \textbf{Ping} & \multicolumn{3}{c|}{5.5~ms} & \multicolumn{3}{c|}{5.5~ms} \\ \hline
    \multicolumn{7}{|c|}{\textbf{HR-5000H Radio}} \\ \hline
    TCP & -- & -- & -- & -- & -- & -- \\ \hline
    UDP test 15~kb/s & 4.924~kb/s & 3.08~\% & 752.57~ms & 5.008~kb/s & 0~\% & 783.53~ms \\ \hline
    UDP test 10~kb/s & 4.792~kb/s & 2.20~\% & 444.37~ms & 4.962~kb/s & 0~\% & 423.52~ms \\ \hline
    UDP test 5~kb/s  & 4.578~kb/s & 0~\%    & 60.63~ms  & 4.574~kb/s & 0~\% & 49.94~ms \\ \hline
    UDP test 4~kb/s  & 3.894~kb/s & 0~\%    & 55.43~ms  & 3.878~kb/s & 0~\% & 64.35~ms \\ \hline
    \textbf{Ping} & \multicolumn{3}{c|}{968~ms} & \multicolumn{3}{c|}{839.75~ms} \\ \hline
    \end{tabular}

    \label{tab:iperf_results_bidirectional}
\end{table*}

\subsection{Session Establishment Performance}
Connection establishment was measured using a centralized timestamp server. Each node sent event notifications over a temporary socket, then the server subtracted start and end timestamps to compute the duration. Seven runs were performed per scenario, and mean, variance and standard deviation were reported. Two scenarios were tested:

\begin{itemize}
    \item \textbf{All‑Wi‑Fi} – all links between nodes used the WLAN (Fig. \ref{fig:Testing_Architecture} labeled 1-4).
    \item \textbf{Radio Backhaul} – the command links (TC1–TC2, TC1–GCS and TC2-GCS) used HR‑5000H radios, while the GCS–UxV link remained on Wi‑Fi (Fig. \ref{fig:Testing_Architecture} labeled 5-8).
\end{itemize}

In direct TC1–GCS connections, Wi‑Fi achieved a mean establishment time of approximately 298 ms with low variance, whereas the radio link averaged 16.1s with a standard deviation exceeding 1s. TC1–TC2 pairing followed the same trend: 195 ms over Wi‑Fi versus 16.5s over the radio. Multi‑hop configurations increased latency further. When TC1 connected to GCS via TC2, the multi‑hop path averaged 0.57s under Wi‑Fi but 18.7s when both hops traversed radios. Adding the UxV: in the TC1-GCS-UxV configuration, if the setup was full Wi-Fi, the total connection was established at 0.275s, while in the radio backhaul, the interval rose to 1.19s. In the TC1-TC2-GCS-UxV configuration, if the setup was full Wi-Fi, the total connection was established at 0.328s, while in the radio backhaul, the interval rose to 1.9s.

These results show that session establishment is dominated by the physical medium. Over Wi‑Fi, the session establishment protocol completes in a few hundred milliseconds, while over HR‑5000H, the high RTT of the waveform, combined with potential packet fragmentation, causes connection times to grow by roughly an order of magnitude.

\subsection{Handover Mechanisms}
\begin{figure}
    \centering
    
    \includegraphics[width=1\linewidth]{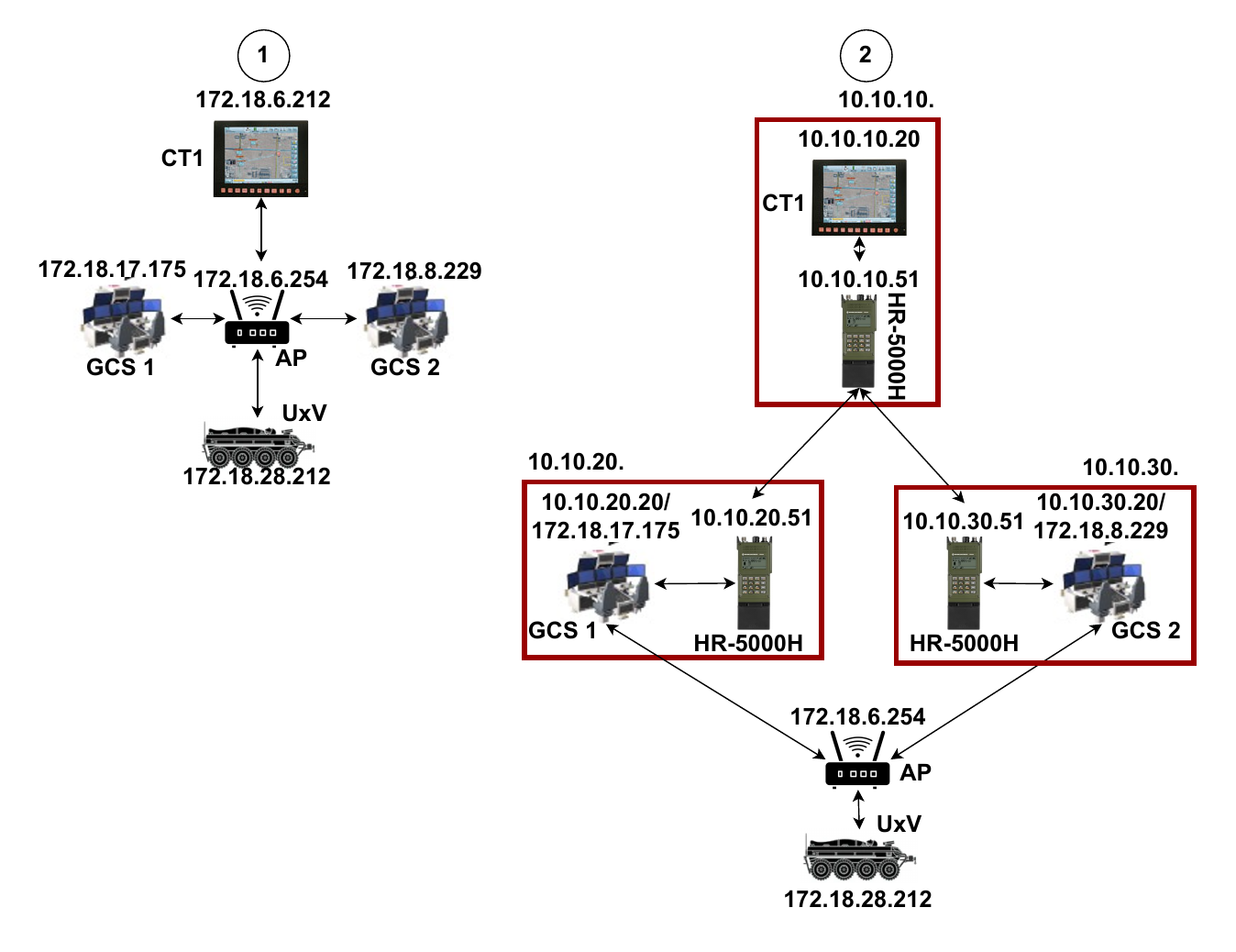}
    \caption{Handover Testing Architecture.}
    \label{fig:Handover_Testing_Architecture}
\end{figure}

Two handover methods were tested with a prototype comprising TC1, two GCSs (GCS1 and GCS2), and a UxV (Fig. \ref{fig:Handover_Testing_Architecture}). Each experiment was repeated seven times using the same centralized timestamp server method explained before.

\subsubsection{Handover with Credential Revocation}

In this method, the operator first revokes the current GCS–UxV credential and then establishes a new session. Because these phases are independent and can partially overlap, two bounds have to be taken into count. The system‑level handover interval, \(\big[\max(\overline{T}_{\text{rev}},\,\overline{T}_{\text{conn}}),\;\overline{T}_{\text{rev}}+\overline{T}_{\text{conn}}\big]\), being the \(\overline{T}_{\text{rev}}\) and the \(\overline{T}_{\text{conn}}\) the mean valeus for the revocation time measure from the time the credRL was sent by the TC1 and the time the UxV received it and shutted down the link to the GCS and the mean value for the connection time from the TC1 to the UxV measure in the previous test.

A more realistic end‑to‑end bound adds the operator’s action latency \(\Delta_{\text{human}}\) thus being \(\big[\max(\overline{T}_{\text{rev}},\,\overline{T}_{\text{conn}})+\Delta_{\text{human}},\;\overline{T}_{\text{rev}}+\overline{T}_{\text{conn}}+\Delta_{\text{human}}\big]\). However, and since this \(\Delta_{\text{human}}\) depends on the operator efficiency, this was not evalueated. With this, in Wi‑Fi, the measured revocation and connection intervals yielded a system‑level bound of 274.6–304.0 ms. When the TC1-GCS link used HR‑5000H radios, the bound expanded to 1.20–1.82s, roughly four to six times higher. Because the GCS–UxV leg remained on Wi‑Fi, this increase originates almost entirely from the high round‑trip time and jitter of the tactical radio.

\subsubsection{Handover with Capacity‑String Modification}

The second method transfers control by forwarding updated credentials. Again, two bounds were computed. Let $\overline{T}_1$ and $\overline{T}_2$ denote the mean credential-update
times for the TC1–GCS1–UxV and TC1–GCS2–UxV chains, respectively.
The system-level bound is
\([\max(\overline{T}_1,\overline{T}_2);\;\overline{T}_1+\overline{T}_2]\). Again, and since the actions are triggered sequentially \(\Delta_{\text{human}}\) as also to be taken into account, although it was not measured. Under Wi‑Fi, the system‑level handover bounds were 49.4–86.9 ms, whereas the HR‑5000H radio raised them to 0.97–1.84s, a \(\approx 20\)× increase. As before, the GCS–UxV leg on Wi‑Fi remained fast. The slowdown is attributed to the TC1-GCS radio segment and the HR‑5000H’s high latency.

\subsection{Key Renewal Performance}

Key renewal tests were run on a three‑node chain (TC1–GCS–UxV). Two configurations were used: Scenario 1, where all links ran over Wi‑Fi (Fig. \ref{fig:Testing_Architecture} label as 2), and Scenario 2, where the TC1–GCS leg used R\&S HR‑5000H radios while the GCS–UxV leg remained on Wi‑Fi (Fig. \ref{fig:Testing_Architecture} label as 6). In both cases, the experiment measured the one‑way message delays and the time each peer took to adopt the new keys.

On Wi‑Fi, one‑way 0x11 transmissions averaged 8 ms, and the GCS and TC1 completed renewal in \(\approx64\) ms and \(\approx38\) ms, respectively. Because the server only activates the new keys after it authenticates a client message with an HMAC computed with the new keys, its measured completion time is longer and depends on the rate at which the client sends data. Substituting the TC1–GCS link with the HR‑5000H radios raised one‑way delays to \(\approx0.4s\) and stretched renewal to \(\approx1.49s\) at the GCS and \(\approx0.86s\) at TC1. A similar pattern emerged on the GCS–UxV leg: Wi‑Fi transmissions were 3–5 ms, but the UxV completed renewal in \(\approx287\) ms versus \(\approx10\) ms at the GCS. 

Overall, the experiments show three trends: (i) 0x11 message propagation is symmetric across uplink and downlink; (ii) server‑side renewal always lags behind the client because it waits for the next authenticated message; and (iii) renewal completion scales with the link’s round‑trip time and the client’s message rate. The TC1–GCS renewal was shorter than the GCS–UxV renewal in the all Wi‑Fi case because TC1 transmits data to the GCS more frequently than the GCS transmits to the UxV.

\subsection{Reliability, Goodput and CPU Performance}

These experiments assessed the robustness and efficiency of NC2S communications under the same TC1–GCS–UxV topology used in the previous tests. Two setups were evaluated: (i) Scenario 1, fully over Wi-Fi; and (ii) Scenario 2, where the TC1–GCS link operated through HR-5000H radios while GCS–UxV remained on Wi-Fi. The test procedure began by establishing a connection between the Tactical Commander (TC1) and the GCS, running it for 30s and then commanding the GCS to connect to the UxV. After two minutes all scripts were terminated, and message counts, byte counts and per‑message CPU times were logged. Because the TNW50 waveform supports only ~10kb/s maximum in Robust Mode and in order to avoid saturating the radio backhaul, the GCS forwarded only MAVLink HEARTBEAT and GPS messages to TC1 every 10s. Also, the HR-5000H manufacturer does not provide the over-the-air datagram structure, making it impossible to determine the complete frame overhead and hence the real physical throughput. Consequently, only the NC2S application-layer goodput could be estimated. Each scenario was executed five times to compute mean, variance and standard deviation.

\subsubsection{Reliability}
In scenario 1, message delivery was almost lossless. Packet loss on the TC1-GCS channels stayed below 0.2\%, and the GCS-UxV channel experienced no losses. Telemetry traffic from the UxV to the GCS, which operates at a higher rate, showed a modest 0.37\% mean loss. These negligible losses and the low dispersion agree with iperf3 benchmarks, where throughput exceeded 50 Mb/s and jitter remained below 1.2 ms. When the HR‑5000H radios were inserted, packet loss increased slightly, with a mean loss being 1.29\% on the TC1-GCS direction, and the remaining directions staying at or below 0.31\%. Even under the TNW50 waveform constraints, the protocol tolerated these minor UDP drops thanks to its stateless design and lack of retransmissions. Overall, both infrastructures delivered highly reliable exchanges.

\subsubsection{Goodput Estimation}

Over Wi‑Fi, the mean application‑layer goodput did not exceed 247.83 bit/s on the \(TC1\rightarrow GCS\) link. The reciprocal \(GCS\leftarrow TC1\) direction required only 134.63 bit/s. Even with the radio backhaul, the goodput remained of the same order (\(\approx250 b/s\)), which is two orders of magnitude below the ~5 kb/s goodput measured for the HR‑5000H using iperf3. This confirms that the HR‑5000H can reliably transport command and credential‑management messages without saturating its channel. In contrast, the UxV-GCS leg used for telemetry demands much higher rates: the \(UxV\rightarrow GCS\) direction averaged \(\approx32 kb/s\), about 100× higher than command traffic. This exceeds the measured radio throughput and explains why continuous UxV control and full telemetry must remain on high‑bandwidth Wi‑Fi. The low variances reported for both scenarios show that NC2S traffic is stable and predictable across runs.

\subsubsection{CPU Processing Time}

Per‑message processing was measured on TC1, GCS and UxV nodes. Under Wi‑Fi, the mean processing times were 2.62 ms for TC1, 2.69 ms for the GCS and 1.02 ms for the UxV, with sub‑millisecond dispersion. When the HR‑5000H backhaul was used, processing delays remained nearly identical with 2.48 ms, 2.59 ms and 1.16 ms respectively. The slightly higher variance on TC1 is attributed to occasional packet bursts caused by the radio’s long latency. Processing therefore contributes \textless 0.1 \% to total end-to-end delay, showing that NC2S performance is network-bound rather than compute-bound.

\subsubsection{Discussion}
The reliability, goodput and CPU measurements confirm that NC2S operates reliably over both Wi‑Fi and tactical radios. With appropriate filtering, command‑and‑control exchanges consume only a few hundred bits per second and fit comfortably within the limited throughput of the HR‑5000H. However, UxV telemetry flows at tens of kilobits per second, exceeding the radio’s capacity. Thus, the HR‑5000H should be reserved for secure command‑node coordination and control handover, while UxV control and full telemetry remain on high‑bandwidth links. The negligible CPU overhead further shows that the system is network‑bound rather than compute‑bound, validating the design choice of lightweight cryptography and stateless UDP messaging for resource‑constrained unmanned vehicles.

\section{Conclusions}

This paper proposes a lightweight and secure C3 architecture that focuses on the integration of commercial UxVs into legacy military C4I systems. It ensures CIA while allowing flexible control delegation between GCSs and UxVs. Existing protocols either impose excessive computational overhead or do not provide sufficient resilience against credential compromise. To address this gap, the NC2S introduces layered authentication based on digital certificates and dynamic credentials, guided by a mission policy defining each node’s privileges and allowed message types.

Extensive laboratory experiments compared NC2S performance over Wi‑Fi and HR‑5000H tactical radios. Tests measured connection establishment, control handover, key renewal, packet loss, goodput and CPU processing time. Under Wi‑Fi, connection establishment, credential renewal and handover completed in milliseconds or sub‑seconds, offering near real‑time responsiveness suitable for UxV control. HR‑5000H links introduced latencies roughly two orders of magnitude higher due to narrow‑band waveforms and internal link mechanisms. Nevertheless, HR‑5000H provided stable communication with minimal packet loss. Cryptographic operations and mTLS‑based authentication added negligible overhead compared with transport delays. Overall, the experimental campaign validated NC2S as a secure, flexible and lightweight framework for UxV control and command delegation, contributing to the development of secure C3 architectures.

Future work should aim to enhance scalability, interoperability, and security. First, the current centralized trust architecture should evolve into a hierarchical and distributed certification model where secondary commanders (TC2) or other trusted entities can issue and sign credentials within mission‑defined boundaries. This would eliminate the single point of failure associated with dependence on a primary commander and improve field autonomy. Second, evaluating NC2S over private 5G networks could extend the GCS–UxV link beyond the short ranges possible with Wi‑Fi. Interoperability with existing Portuguese C4I systems such as the BMS and the TAK should also be pursued to enable seamless integration, enhanced situational awareness and joint mission coordination. Additional security enhancements include incorporating perfect forward secrecy, active defenses against denial‑of‑service attacks, and an internal encryption layer to guarantee end‑to‑end confidentiality over open channels. Finally, large‑scale and hardware‑in‑the‑loop testing campaigns with multiple commanders and real UxV platforms are essential to evaluate scalability, latency and mission reliability under realistic conditions.

\section*{Acknowledgments}

This work was developed within the scope of the Portuguese Army’s Project EXE02 – Remote and Autonomous Systems, specifically in EXE03 – Unmanned Ground Systems, through the eMOVE – Robotisation of the M113 initiative. The work of António Grilo and Carlos Ribeiro was funded by national funds through FCT — Fundação para a Ciência e a Tecnologia, I.P., under projects/supports UID/6486/2025 (\url{https://doi.org/10.54499/UID/06486/2025}) and UID/PRR/6486/2025 (\url{https://doi.org/10.54499/UID/PRR/06486/2025}).

The authors would like to thank Lieutenant-Colonel Pena Madeira of the Military Academy, as well as the SIC-T team of the Portuguese Army, who provided the HR-5000H radios and collaborated for three consecutive weeks during the experimental testing of the protocol. Their availability, expertise, and generosity were crucial to the obtained results.

\vspace{12pt}
\begin{IEEEbiography}[{\includegraphics[width=1in,height=1.25in,clip,keepaspectratio]{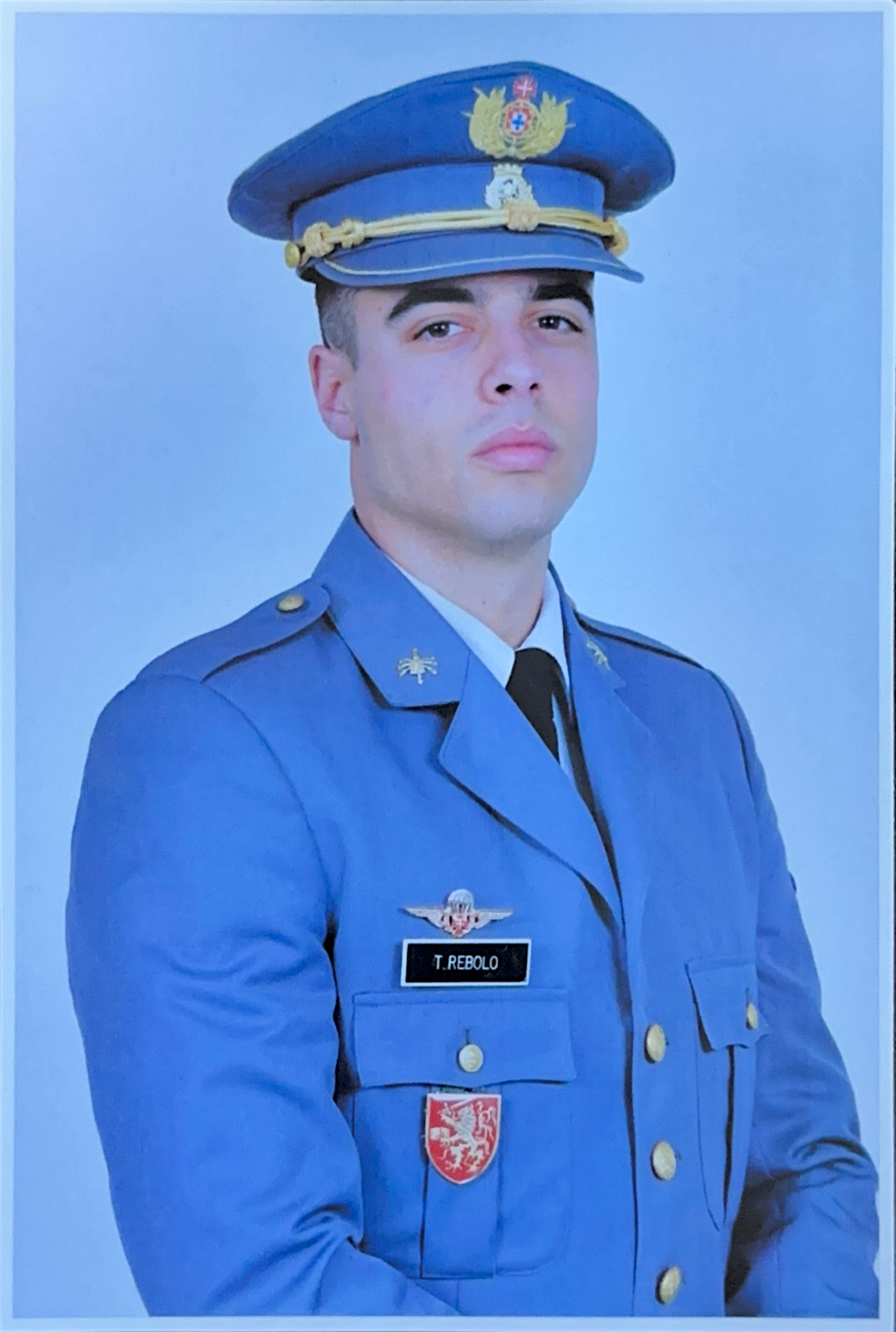}}]{Second-Lieutenant (OF-1) Tomás Rebolo
} Tomás Rebolo received the B.Sc. degree in Military Electrical and Computers Engineering from the Portuguese Military Academy, Lisbon, Portugal, in 2023. He serves as an Officer in the Portuguese Army within the Signals branch. His professional and research interests include artificial intelligence models, communication protocols, network management and optimization.
\end{IEEEbiography}

\begin{IEEEbiography}[{\includegraphics[width=1in,height=1.25in,clip,keepaspectratio]{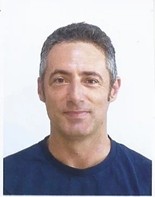}}]{Prof. António Grilo} holds the Ph.D. degree in Electrical and Computer Engineering from IST, where he is currently Associate Professor. At IST, he teaches courses related with Mobile Communications, IoT and Smart Grids. Since 1996, he has been working in European Commission (EC) projects related with communication networks. He is currently a Researcher with the Intelligent Communication Networks group of INESC INOV-Lab, in Lisbon. He is the author or the co-author of more than eighty scientific articles on subjects related with communication networks. His current research interests include UxV networks, Internet of Things, Edge Computing optimization, and applications of Artificial Intelligence in communication systems and networks. He is Senior Member of IEEE and Life Member of AFCEA.
\end{IEEEbiography}


\begin{IEEEbiography}
[{\includegraphics[width=1in,height=1.25in,clip,keepaspectratio]{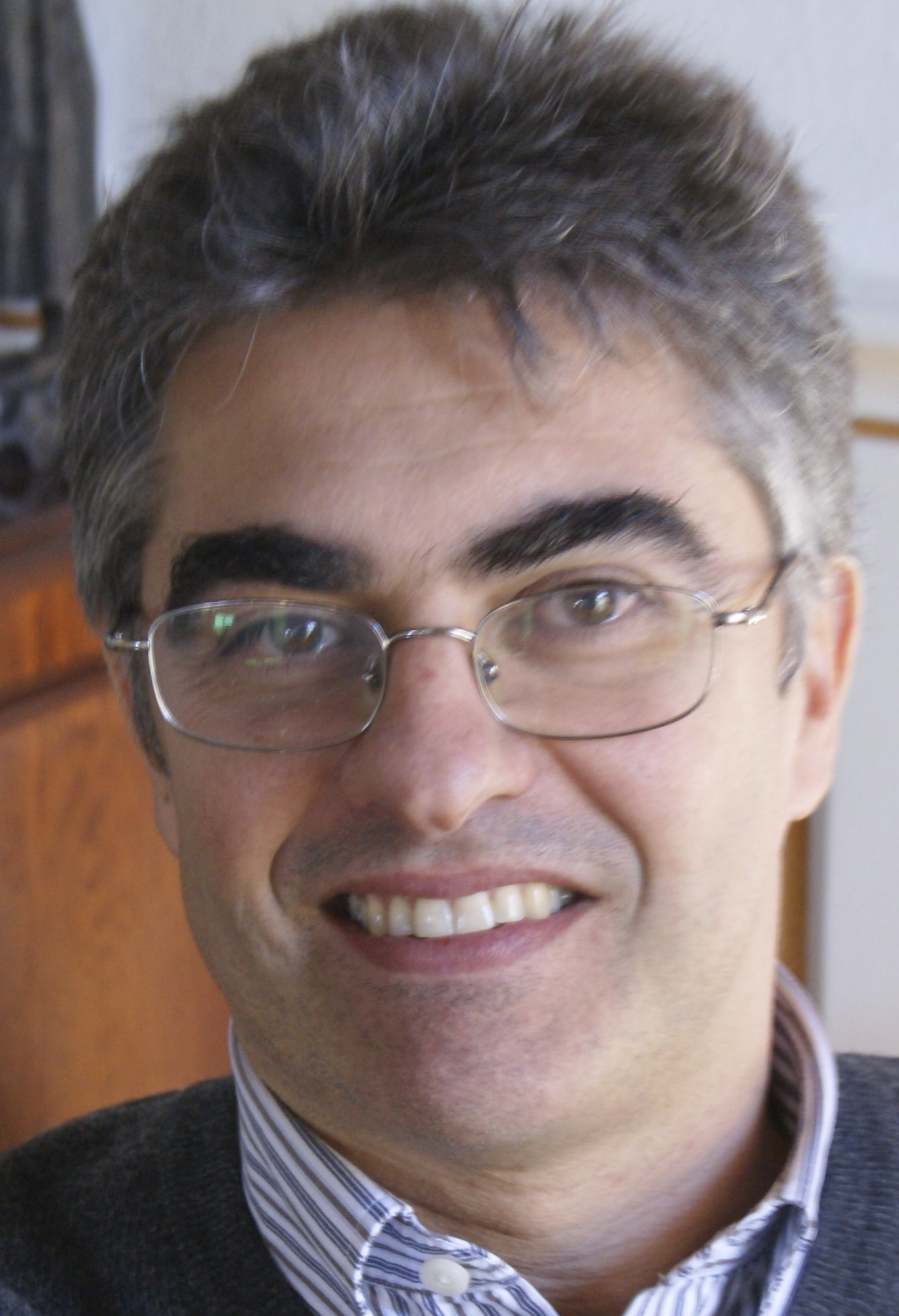}}]{Prof. Carlos Ribeiro} has been a professor at Instituto Superior Técnico, University of Lisbon, for over 30 years and was vice-rector of the University of Lisbon for 10 years. He is currently a researcher in the cybersecurity unit of the INOV research center in Lisbon.  His main area of research has been computer security, particularly authentication solutions in many different environments, from large-scale web identity systems to sensor networks, with different requirements, like e-voting or e-money, in which he has published and participated in several research projects.
\end{IEEEbiography}
\end{document}